\documentclass{LT23auth}
\usepackage{graphicx}

\begin{document}

\begin{frontmatter}

% Use lower case letters in the title.
\title{Bose-Fermi mixed condensates of atomic gas 
with Boson-Fermion quasi-bound state}

\author[address1]{Hiroyuki Yabu \thanksref{thank1}},
\author[address2]{Yashutoshi Takayama},
\author[address1]{Toru Suzuki}

\address[address1]{Department of Physics, Tokyo Metropolitan University, 
1-1 Minami-Ohsawa, Hachioji, Tokyo 192-0397, Japan}

\address[address2]{KEK (High energy accelerator research organization), 
                   1-1 Oho, Tsukuba, Ibaraki 305-0801 Japan}

% The corresponding author should be distinguished and his email
% address and/or fax number must be given. His mailing address has to
% be complete: the proofs are send to this address around
% January 1, 2003. The address for sending proofs has to be indicated
% as "present address", if it is different from the address above.
\thanks[thank1]{ E-mail:yabu@comp.metro-u.ac.jp}

\begin{abstract}
The phase structures of the boson-fermion (B and F) 
mixed condensates 
of atomic gas are discussed 
under the existence of 
boson-fermion composite fermions (quasi-bound states) BF
from the equilibrium in 
$\mathrm{B}+\mathrm{F} \leftrightarrow \mathrm{BF}$. 
Especially we discuss the competitions 
between the BF degenerate states 
and the Bose-Einstein condensates (BEC) in low-T. 
The criterion for the BEC realization 
is obtained from the algebraically-derived 
phase diagrams at $T=0$.
\end{abstract}

%
% write here 3 or 4 keywords separated by semicolons
%
\end{frontmatter}

\section{Introduction}

Experimental successes of the BEC and the fermi degenerate systems
of the trapped atomic gas have opened up renewed interests 
in the boson-fermion mixed condensates, 
which are expected to show many interesting physical 
phenomena \cite{TMU}. 

In case that the boson-fermion interaction is enough attractive, 
the boson-fermion pairs can make quasi-bound states,  
which behave as composite fermions BF, 
and produce new phases as the BF degenerate state. 

In this paper, 
we discuss the phase structures of the mixed condensates
under the existence of the quasi-bound states 
with solving the equilibrium condition 
for the reaction: 
$\mathrm{B} +\mathrm{F} \leftrightarrow \mathrm{BF}$. 
Especially interesting is a competition 
between the BF degenerate states and the BEC 
of unpaired bosons; 
the energy-reduction in the BF binding energies 
v.s. that in the boson kinetic energies in the BEC. 
If the BF binding energy is enough large, 
the BF pairs can exhaust the bosons 
and the BEC will not appear in the mixed condensates.

\section{Equilibrium condition}

Let's consider the uniform system of porallized bosons 
and fermions (B and F), 
with the masses $m_{\mathrm{B}}$ and $m_{\mathrm{F}}$. 
The total numbers of B and F should be conserved 
and their densities are $n_{\mathrm{Btot}}$ 
and $n_{\mathrm{Ftot}}$ each other. 
A quasi-bound state (composite fermion) BF is assumed to 
exist with the mass $m_{\mathrm{BF}}$. 

In states with temperature $T$, 
because of the equilibrium 
$B+F \leftrightarrow BF$,
a part of the atom B and F are paired in the BF states, 
and the others are in free unpaired states.
The equilibrium condition is given by 
\begin{equation}
     \mu_{\mathrm{B}} +\mu_{\mathrm{F}} 
     =\mu_{\mathrm{BF}} +\Delta{m} c^2,
\label{eQa}
\end{equation}
where $\mu_{\mathrm{B,F,BF}}$ are chemical potentials 
of atoms B, F, BF each other, 
and $\Delta{m} =m_{\mathrm{BF}} -m_{\mathrm{B}} -m_{\mathrm{F}}$ 
is a binding energy of the BF state. 
The chemical potentials in (\ref{eQa}) are obtained by 
the density formulae of the free bose/fermi gas:
\begin{eqnarray}
     n_{\mathrm{B}} &=& \frac{(m_{\mathrm{B}})^{3/2}}{\sqrt{2} \pi^2}
          \int_0^\infty\frac{\sqrt{\epsilon}d\epsilon}{
               e^{(\epsilon-\mu_{\mathrm{B}})/k_B T}-1}, 
      \label{eQb}\\
     n_a &=& \frac{(m_a)^{3/2}}{\sqrt{2} \pi^2}
          \int_0^\infty\frac{\sqrt{\epsilon}d\epsilon}{
               e^{(\epsilon-\mu_a)/k_B T}+1}, 
     \quad (a =\mathrm{F},\mathrm{BF}) 
      \label{eQd}
%
%     n_{\mathrm{BF}} &=& \frac{(m_{\mathrm{BF}})^{3/2}}{\sqrt{2} \pi^2}
%          \int_0^\infty\frac{\sqrt{\epsilon}d\epsilon}{
%               e^{(\epsilon-\mu_{\mathrm{BF}})/k_B T}+1}, 
%      \label{eQd}
\end{eqnarray}
where  $k_B$ is a Boltzmann constant, 
and $n_{\mathrm{B,F,BF}}$ are the densities of 
the free (unpaired) B and F and the composite BF. 

Solving eq.~(\ref{eQa}) with (\ref{eQb}-\ref{eQd}) 
under the atom number conservation for B and F: 
$n_{\mathrm{B}} +n_{\mathrm{BF}} =n_{\mathrm{Btot}}$ and
$n_{\mathrm{F}} +n_{\mathrm{BF}} =n_{\mathrm{Ftot}}$,  
we obtain the densities $n_{\mathrm{B,F,BF}}$ 
as functions of $T$ and $n_{\mathrm{Btot},\mathrm{Ftot}}$.

When $T$ and $n_{\mathrm{B}}$ satisfy 
$T < T_C \equiv \frac{2\pi \hbar^2}{m_B k_B} 
\left(\frac{n_B}{2.613}\right)^{2/3}$, 
a part of free bosons condensates into the BEC, 
and $\mu_{\mathrm{B}}$ becomes zero.
In that case, the equilibrium condition 
becomes 
$\mu_{\mathrm{F}} 
=\mu_{\mathrm{BF}} +\Delta{m} c^2$.
When the BEC exists, 
The condensed- and normal-component densities 
of bosons $n_{\mathrm{BEC,Bnor}}$ 
are defined by 
$n_{\mathrm{BEC}} =n_{\mathrm{B}}
          \left[ 1 -\left(\frac{T}{T_C}\right)^{3/2} \right]$, 
and  
$n_{\mathrm{Bnor}} =n_{\mathrm{B}}-n_{\mathrm{BEC}}$.

\section{Results and summary}

\begin{figure}[tbp]
%h=here, t=top, b=bottom, p=separate figure page
\begin{center}\leavevmode
\includegraphics[width=0.9\linewidth]{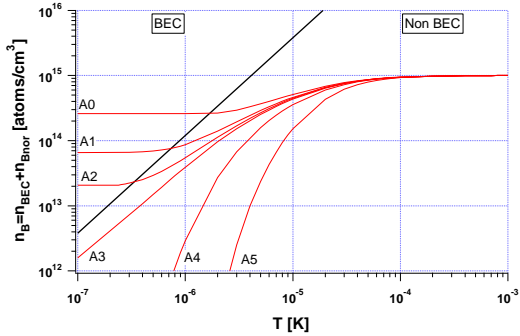}
\caption{$T$-dependence of boson density}
\label{fig1}\end{center}\end{figure}
As an typical example, 
we show the $T$-dependence of free boson density 
$n_{\mathrm{B}} =n_{\mathrm{BEC}}+n_{\mathrm{Bnor}}$ 
when $n_{\mathrm{Btot}} =n_{\mathrm{Ftot}} 
=10^{15}\,{\rm atoms/cm^3}$ in Fig.~1. 
The lines A0-A5 are for 
$\Delta{m}=(0,-3,-4,-4.71,-10) \times 10^{-6}\,{\rm K}$. 
The oblique straight line is the critical border 
of the BEC region.
The $n_{\mathrm{B}}$ are found to decrease with decreasing $T$; 
it is because the number of composite fermions increases 
in low-$T$.
In small $\Delta{m}$ cases (A0-A3), 
the $n_{\mathrm{B}}$ is still large and 
free bosons can condensate into the BEC in low-$T$, 
but, in large $\Delta{m}$ cases (A4,A5), 
free bosons are exhausted in making composite fermions 
and the $n_{\mathrm{B}}$ becomes too small 
for the BEC realization.
The line A6 corresponds to the critical case. 
In high-$T$ region, all composite fermions dissociate 
into free bosons and fermions, 
so that $n_{\mathrm{B}}$ approaches to $n_{\mathrm{Btop}}$.  
\begin{figure}[btp]
%h=here, t=top, b=bottom, p=separate figure page
\begin{center}\leavevmode
\includegraphics[width=0.9\linewidth]{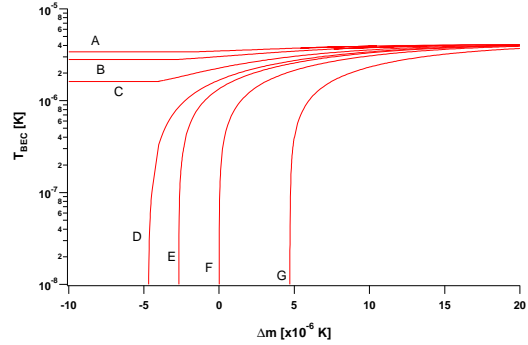}
\caption{$T_C$ for BEC transition v.s. $\Delta{m}$}
\label{fig3}\end{center}\end{figure}
In Fig.~2, the variations of $T_C$ for the BEC 
transitions are shown 
in $n_{\mathrm{F}}/n_{\mathrm{B}} =$0.3 (A), 
0.5 (B), 0.8 (C), 1 (D), 1.2 (E), 1.36 (F), 2 (G).

When $n_{\mathrm{Btot}} < n_{\mathrm{Ftot}}$, 
the $n_{\mathrm{B}}$ has similar $T$-dependence 
as in Fig.~1. 
When $n_{\mathrm{Btot}} > n_{\mathrm{Ftot}}$, 
the BEC always occur in enough low-$T$
because, after all fermions are paired, 
the free bosons still remain. 
%\section{Phase diagrams at T=0 and BEC criterion}

\begin{figure}[btp]
%h=here, t=top, b=bottom, p=separate figure page
\begin{center}\leavevmode
\includegraphics[width=0.9\linewidth]{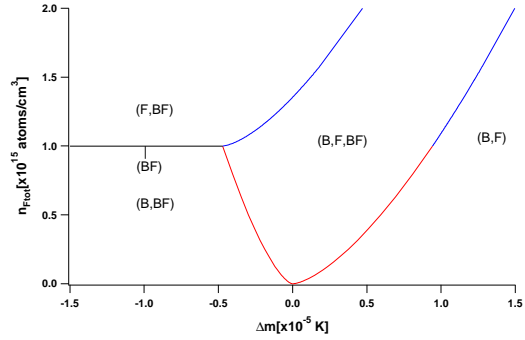}
\caption{Phase diagram at $T=0\,{\rm K}$ 
         ($n_{\mathrm{Btot}} =10^{15}\,{\rm atoms/cm^3}$)}
\label{fig2}\end{center}\end{figure}
At $T=0$, the condition (\ref{eQa}) becomes 
$0+\epsilon_{\mathrm{F}} 
=\epsilon_{\mathrm{BF}} =\Delta{m} c^2$, 
where 
$\epsilon_a =\frac{(3\pi^2)^{2/3}}{2^{1/3} m_a} n_a^{2/3}$ 
($a=B,BF$).
It can be solved algebraically and gives the phase structures
at $T=0$. 
In Fig.~3, we show the phase diagrams 
in $n_{\mathrm{Ftot}}-\Delta{m}$ plane 
when $n_{\mathrm{Ftot}} =10^{15}\,{\rm atoms/cm^3}$, 
where the symbol (B,F,BF) means 
the coexistence of free bosons and 
free and composite fermions,
and so on. 

From this diagram, 
we can read off the criterion for the BEC to occur; 
it should occur in the regions when 
free bosons exist at $T=0$, e.g 
the ones with the symbol B in Fig.~2. 

In summary, we studied the role of the composite fermion 
in the boson-fermion mixed condensates 
and its phase structure in low-$T$. 
The more details and further applications 
of the present results should be discussed  
in further publication\cite{TSYS}.

%
%  Here is the template to include a figure.
%
%
%
% for acknowledgement
%
%\begin{ack}
%
%\end{ack}
%
%
% The format of reference should be
% Author1, Author2, Author3, Journal {\bf volume} (year) page.
% No ``and'' between the authors are necessary. 
%


\begin{thebibliography}{9}
\bibitem{TMU} 
T. Miyakawa, T. Suzuki, H. Yabu, 
Phys.~Rev. {\bf A64} (2001) 033611 
and references therein. 
\bibitem{TSYS}
Y. Takayama, T. Suzuki, H. Yabu, P. Schuck, 
in preparation.
\end{thebibliography}
\end{document}